\documentclass[aps,prd,nofootinbib,showkeys showpacs,preprint,onecolumn]{revtex4-1}
\pdfoutput=1
\usepackage{latexsym}
\usepackage{graphicx}
\usepackage{dcolumn}
\usepackage{bm}
\usepackage{slashed}
\usepackage{amsmath,graphicx}
\usepackage{amssymb}
\usepackage[colorlinks=true,linktocpage=true,linkcolor=blue,citecolor=blue]{hyperref}
\usepackage{float}
\usepackage{nicefrac}
\usepackage[normalem]{ulem}
\usepackage{amsmath}
\usepackage{subfigure} 
\usepackage{bbold} 
\usepackage[makeroom]{cancel}
\usepackage{stmaryrd} 

\def\be{\begin{equation}}
	\def\ee{\end{equation}}
\newcommand{\bel}[1]{\begin{eqnarray}\label{#1}}
	\newcommand{\eel}{\end{eqnarray}}
\def\barr{\begin{array}}
	\def\earr{\end{array}}
\def\beq{\begin{eqnarray}}
	\def\eeq{\end{eqnarray}}
\def\bfig{\begin{figure}}
	\def\efig{\end{figure}}

\newcommand{\nn}{\nonumber}

\newcommand{\sh}[1]{\sinh#1}
\newcommand{\ch}[1]{\cosh#1}

\newcommand{\lab}[1]{\label{#1}}
\def\nn{\nonumber}

\begin{document}
 
    \title{Spin alignment of vector mesons by second-order
 hydrodynamic gradients} 
	\author{Avdhesh Kumar} 
	\email{avdheshk@gate.sinica.edu.tw} 
	\affiliation{Institute of Physics, Academia Sinica, Taipei, 11529, Taiwan}
 \author{Philipp Gubler} 
	\email{philipp.gubler1@gmail.com} 
	\affiliation{Advanced Science Research Center, Japan Atomic Energy Agency,
Tokai, Ibaraki 319-1195, Japan}
	\author{Di-Lun Yang}
\email{dlyang@gate.sinica.edu.tw}
\affiliation{Institute of Physics, Academia Sinica, Taipei, 11529, Taiwan}
	\date{\today} 
	\bigskip
	\begin{abstract}
	Starting with the polarization dependent Wigner function of vector mesons, we derive an expression for the {00}--component ($\rho_{00}$) of spin density matrix in terms of the second order gradients of the vector meson distribution functions. We further apply a thermal model 
 to analyze the transverse momentum and the azimuthal angle dependence of $\rho_{00}$ for $\phi$ and $K^{*0}$ mesons resulting from distribution gradients in Au-Au collisions with $\sqrt{s_{NN}}=130$ GeV at mid-rapidity.
Our results for the transverse momentum dependence indicate that 
 the deviations of $\rho_{00}$ from $1/3$ as the signal for spin alignment are greatly enhanced at large transverse momenta and have a strong centrality dependence while analysis of the azimuthal angle ($\phi_q$) dependence suggest that such deviations have a $\cos(2\phi_q)$ structure with opposite sign for $\phi$ and $K^{*0}$.   
 Our finding may be considered as a baseline for probing spin-alignment mechanisms beyond hydrodynamic gradients.
\end{abstract}

\date{\today}

	
	
\maketitle
%
%
\section{Introduction}
\label{sec1:intro}
Strongly interacting matter created in non-central collisions of two heavy nuclei at relativistic beam energies carries a huge amount of global orbital angular momentum along the direction perpendicular to the reaction plane. A fraction of such orbital angular momentum can get transferred into the spin  degrees of freedom which in turn may lead to the spin polarization of emitted particles~\cite{Liang:2004xn,Voloshin:2004ha,Voloshin:2017kqp,Liang:2004ph,Betz:2007kg,Becattini:2007sr,Gao:2007bc} in a way similar to the Barnett effect~\cite{PhysRev.6.239}
and the Einstein–de Haas effect~\cite{Einstein} in materials. As a matter of fact, a non-zero global and local spin polarization of hyperons has been measured by the STAR Collaboration~\cite{STAR:2017ckg,STAR:2019erd,Adam:2018ivw,STAR:2020xbm} at BNL, the ALICE Collaboration at CERN~\cite{ALICE:2019onw}, and the HADES Collaboration at GSI~\cite{Kornas:2020qzi}. Theoretically, relativistic hydrodynamics that makes use of the global thermodynamic equilibrium formula, which connects the mean spin pseudo-vector of a fermion with the thermal vorticity~\cite{Becattini2013a,Becattini:2016gvu,Karpenko:2016jyx,Becattini:2017gcx,Fang:2016vpj}, turned out be a successful tool in describing the experimentally measured global polarization of $\Lambda$-hyperons  ~\cite{Karpenko:2016jyx,Xie:2017upb,Pang:2016igs,Becattini:2017gcx,Li:2017slc,Wei:2018zfb,Ryu:2021lnx}. Surprisingly, the predictions for the {\it local} spin polarization, i.~e. the momentum dependence of the longitudinal spin polarization\cite{Becattini:2017gcx,Xia:2018tes}, fails to agree with the measured values~\cite{STAR:2019erd}.  
This result has motivated further theoretical developments, trying to clarify the origins of spin polarization and spin transport phenomena in relativistic heavy ion collisions \cite{Hidaka:2017auj,Liu:2020dxg,Liu:2021uhn,Becattini:2021suc,Yi:2021ryh,Hidaka:2017auj,Hidaka:2018ekt,Yang:2018lew,Shi:2020htn,Becattini:2021suc,Buzzegoli:2021wlg,Fang:2022ttm,Wang:2022yli,Lin:2022tma,Weickgenannt:2022zxs,Weickgenannt:2022qvh,Bhadury:2020puc,Bhadury:2020cop,Buzzegoli:2022kqx,Bhadury:2022ulr,Hidaka:2016yjf,Gao:2019znl,Weickgenannt:2019dks,Hattori:2019ahi,Wang:2019moi,Yang:2020hri,Wang:2020pej,Weickgenannt:2020aaf,Fang:2023bbw} (see   references ~\cite{Becattini:2022zvf,Hidaka:2022dmn} for recent reviews). Such developments include investigating the roles of the symmetric gradients of hydrodynamic variables (known as thermal shear)~\cite{Hidaka:2017auj,Liu:2021uhn,Becattini:2021suc}, gradients of chemical potentials~\cite{Hidaka:2017auj,Liu:2020dxg} and spin potentials~\cite{Buzzegoli:2021wlg}.
Interestingly, taking into account thermal shear corrections in local equilibrium, theoretical results for the local spin polarization of $\Lambda$ hyperons agree with the experimental data only if the effect of temperature gradients are neglected~\cite{Becattini:2021iol} or the $\Lambda$ hyperon mass is replaced with it's constituent strange quark mass~\cite{Fu:2021pok,Florkowski:2021xvy}. 

 Unlike hyperons, vector mesons decay via strong or electromagnetic interactions in which parity is conserved. Therefore, the spin polarization of vector mesons can not be measured directly as the direction of polarization is not known. However, the spin alignment of vector mesons can be studied by measuring the deviations from the equilibrium value ($1/3$) of
$\rho_{00}$, which parametrizes the only independent degree of freedom among the three diagonal elements 
 ($\rho_{00}$, $\rho_{11}$, $\rho_{-1,-1}$) of the $3\times3$ hermitian spin density matrix with unit trace, assuming that parity is conserved 
 ~\cite{Schilling:1969um,Park:2022ayr}.
 Experimental Measurements~\cite{ALICE:2019aid,ALICE:2022dyy,Mohanty:2021vbt,STAR:2022fan} indicate that the global spin alignment of vector mesons is much larger than theoretical predictions based on the assumption of thermal equilibrium~\cite{Becattini:2007sr,Becattini:2007nd} and the spin coalescence model~\cite{Liang:2004xn,Yang:2017sdk}. 
  Furthermore, both quantitative and qualitative differences exist  between distinct flavors and different collision energies~\cite{ALICE:2019aid,ALICE:2022dyy,Mohanty:2021vbt,STAR:2022fan}. For example, at LHC energies, $\rho_{00}<1/3$ is observed for both $\phi$ and $K^{*0}$ mesons with small transverse momenta~\cite{ALICE:2019aid} while at RHIC energies, $\rho_{00}>1/3$ for $\phi$ and $\rho_{00}\approx 1/3$ for $K^{*0}$ were globally found ~\cite{STAR:2022fan} (see also  Ref.~\cite{ALICE:2022dyy} for some experimental results  of $J/\psi$ spin alignments). 
Such results have motivated the development of various theoretical mechanisms~\cite{Sheng:2019kmk,Sheng:2020ghv,Xia:2020tyd,Muller:2021hpe,Yang:2021fea,Goncalves:2021ziy,Sheng:2022wsy,Li:2022neh,Sheng:2022ffb,
Li:2022vmb,Wagner:2022gza,Muller:2021hpe,Yang:2021fea,Kumar:2022ylt,Kumar:2023ghs,Sheng:2023urn,DeMoura:2023jzz,Yin:2024dnu,Dong:2023cng} but the issue remains unresolved. 
We note especially that the treatment of hydrodynamic gradients in this context has so far been incomplete. 
Specifically, only a negligible contribution from such 
gradients was estimated based on the small spin polarization from the vorticity of strange quarks in the coalescence model (see e.g. \cite{Xia:2020tyd}). A similar result is expected based on the shear-induced polarization (governed by the thermal shear tensor)~\cite{Fu:2021pok} of strange quarks in the coalescence model.
The goal of this paper is to fill this gap by providing a more rigorous estimation of the effects of hydrodynamic gradients to the spin 
alignment of vector mesons in heavy ion collisions. This will provide a baseline for more exotic mechanisms such as the fluctuating 
background fields stemming from the strong interaction \cite{Sheng:2019kmk,Sheng:2020ghv,Xia:2020tyd,Sheng:2023urn,Muller:2021hpe,
Yang:2021fea,Kumar:2022ylt, Kumar:2023ghs} (see also Refs.~\cite{Ma:2018qvc,Stebel:2021bbn,Hauksson:2023tze} for related studies in proton-proton and proton-nucleus collisions and jet polarization).

In this paper, starting with the Wigner function of vector mesons and its expansion up to the second order in $\hbar$, equivalent to second order space-time gradients, we derive an expression for the 00-component of the spin density matrix ($\rho_{00}$) in terms of the distribution functions of vector mesons, 
for which the primary results are shown in Eqs.~(\ref{eq:rho00q})-(\ref{eq:rho_fac_deno}). We further discuss the high momentum and low momentum limits of $\rho_{00}$ as shown in Eq.~(\ref{eq:rho001}) and Eq.~(\ref{eq:rho002}), respectively. In order to quantitatively estimate the order of magnitude of $\rho_{00}$ from such contributions, we adopt a thermal model with single-freeze-out \cite{Broniowski:2001we} to evaluate $\rho_{00}$ for $\phi$ and $K^{*0}-$mesons. 

The paper is structured as follows. In Sec.~\ref{Sec:wf}, we derive the Wigner functions of polarized vector mesons up to order $\hbar^2$. In Sec.~\ref{sec:SPD}, we  derive a new equation for the $\rho_{00}$-component of the spin density matrix in terms of the gradients of vector meson distributions and provide simplified expressions in the high and low momentum limits. In Sec.~\ref{sec:thm}, we outline the main steps involved in the calculation of various experimental observables related to spin alignments of vector meson using the thermal model with single freeze-out. In Sec.~\ref{sec:res}, we discuss our numerical results. In Sec.~\ref{sec:summary_outlook}, we present a summary of this work and an outlook. Some technical details are provided in the appendices.

{\it Notation and conventions:} 
Throughout this paper 
the mostly minus signature of the Minkowski metric $\eta^{\mu\nu} = {\rm diag} (1, -1,-1,-1)$ is used. The notations $A^{(\mu}B^{\nu)}\equiv A^{\mu}B^{\nu}+A^{\nu}B^{\mu}$ and $A^{[\mu}B^{\nu]}\equiv A^{\mu}B^{\nu}-A^{\nu}B^{\mu}$ are employed to define symmetric and antisymmetric tensors respectively. Greek letters ($\mu, \nu$ etc) are used for space and time components (indices take on values $0, 1, 2, 3$) while Latin letters $i,j$ etc stand for spatial components (taking on values $1, 2, 3$). 
   
\section{Wigner functions of vector mesons} \label{Sec:wf}
The spin-1 vector-meson field can be 
expanded as~\cite{ Sheng:2022ffb},
\begin{eqnarray}
	V^{\mu}(x)=\sum_{\lambda=\pm 1,0}\int \frac{d^3k}{(2\pi)^3\sqrt{2E_k}}\big[\epsilon^{\mu}(\lambda, k)a(\lambda,\bm k)e^{-ik\cdot x}+\epsilon^{*\mu}(\lambda, k)b^{\dagger}(\lambda,\bm k)e^{ik\cdot x}\big],
\label{mode_exp}
\end{eqnarray}
where $k\equiv{(E_k,\bm k)}$ is the on-shell four-momentum with $E_k=\sqrt{|\bm k|^2+M^2}$ being the energy, $\bm k$ the momentum and $M$ the mass of the considered vector meson. $a^{\dagger}(\lambda,\bm k)$ and $a(\lambda,\bm k)$ are the creation and annihilation operators for particles while $b^{\dagger}(\lambda,\bm k)$ and $b(\lambda,\bm k)$ are those for anti-particles, which follow the respective mesonic commutation relations. 
In principle, for interacting vector mesons, there could be further corrections to the wave functions beyond the simple 
plane waves in the above mode expansion. For our consideration, we focus on the simplest case by ignoring such corrections and instead absorbing the interaction-dependent corrections into the creation and annihilation operators or more precisely the distribution functions, in which the interaction dependence is dictated by the kinetic equation after the Wigner transformation. 
The four vector $\epsilon^{\mu}(\lambda, k)$ is the polarization vector. Following Refs.~\cite{Sheng:2022ffb,Kumar:2023ghs}, it can for a vector meson be written as 
\begin{eqnarray}
\epsilon^{\mu}(\lambda, k)&=&\Big(\frac{-k_{\alpha}\epsilon^{\alpha}_{{\lambda},\perp}}{M},\epsilon^{\mu}_{{\lambda},\perp}-\frac{k_{\alpha}\epsilon^{\alpha}_{{\lambda},\perp}}{M(E_q+M)}k^{\mu}_{\perp}\Big), \quad\quad\quad \label{pol_vec1}
\end{eqnarray}
where $V^{\mu}_{\perp}=\Delta^{\mu\nu}V_{\nu}$ with $\Delta^{\mu\nu}=\eta^{\mu\nu}-n^{\mu}n^{\nu}$ and $n^{\mu}=(1,\bm 0)$. The four-vector ${\bm\epsilon^\alpha_{\lambda,\perp}}\equiv{(\epsilon^0_{\lambda,\perp},\bm\epsilon^i_{\lambda,\perp})}$ satisfies the condition ${\epsilon^0_{\lambda,\perp}}=0$ and reduces to  ${(0,\bm\epsilon_{\lambda})}$.
The three vector $\bm\epsilon_{\lambda}$ is the spin-state vector, which depends on the spin quantization axis and satisfies the following relation 
\begin{equation}
\bm\epsilon_{\lambda}\cdot \bm\epsilon^{*}_{\lambda'}=\delta_{\lambda\lambda'}. \label{eps_r1}
\end{equation}
It can be shown that $\epsilon^{\mu}(\lambda,k)$ and $\epsilon^{*}_{\mu}(\lambda',k)$ satisfy 
\begin{equation} 
\epsilon^{\mu}(\lambda,k)\epsilon^{*}_{\mu}(\lambda',k)=-\bm\epsilon_{\lambda}\cdot \bm\epsilon^{*}_{\lambda'}=-\delta_{\lambda\lambda'},
\end{equation}
and
\begin{equation}
\epsilon^{\mu}(\lambda,k)k_{\mu}=0.
\end{equation}
In the vector meson rest frame, we have $\epsilon^{\mu}(\lambda,0)=(0,\bm\epsilon_{\lambda})$.

The Wigner function for vector mesons in phase space can be written as \cite{Hattori:2020gqh}  
\begin{eqnarray} 
&&W^{<\mu\nu}(q,X) \\ \nonumber
 &=&\int d^4Ye^{iq\cdot Y}\langle V^{\dagger\nu}(X-Y/2)V^{\mu}(X+Y/2)\rangle
	\\\nonumber
	&=&\pi\sum_{\lambda,\lambda'=\pm 1,0}\int\frac{d^3k_{-}}{(2\pi)^3}\frac{e^{-ik_-\cdot X}}{\Big[\Big(|\bm q|^2+\frac{|\bm k_-|^2}{4}\Big)^2-(\bm q\cdot \bm k_{-})^2+2M^2\Big(|\bm q|^2+\frac{|\bm k_-|^2}{4}\Big)+M^4\Big]^{1/4}}
	\\\nonumber	&&\times\bigg[\epsilon^{\mu}\Big(\lambda, q+\frac{k_-}{2}\Big)\epsilon^{*\nu}\Big(\lambda', q-\frac{k_-}{2}\Big)\Big\langle a^{\dagger}\Big(\lambda', \bm q-\frac{\bm k_-}{2}\Big)a\Big(\lambda, \bm q+\frac{\bm k_-}{2}\Big)\Big\rangle\delta(q^0-k_+^0)
	\\\nonumber
	&&+\epsilon^{\nu}\Big(\lambda', -q+\frac{k_-}{2}\Big)\epsilon^{*\mu}\Big(\lambda, -q-\frac{k_-}{2}\Big)\Big\langle b\Big(\lambda', -\bm q+\frac{\bm k_-}{2}\Big)b^{\dagger}\Big(\lambda, -\bm q-\frac{\bm k_-}{2}\Big)\Big\rangle\delta(q^0+k_+^0)\bigg]\label{eq:Wigner_V1}
	,
\end{eqnarray}
where
\begin{eqnarray}
	k_+^0=\frac{1}{2}\Big(E_{{q}+\frac{k_-}{2}}+E_{{q}-\frac{k_-}{2}}\Big)
	\, , \qquad 
	k_-^0=\Big(E_{{q}+\frac{k_-}{2}}-E_{{q}-\frac{k_-}{2}}\Big).
\end{eqnarray} 
For $\phi$ mesons, $a(\lambda,\bm k) = b(\lambda,\bm k)$. 
In order to carry out the integration over $ k_-$, 
we expand the integrand for small $k_-$ and keep the terms up to $\mathcal{O}(k^2_-)$, which corresponds to the $\hbar$ expansion up to $\mathcal{O}(\hbar^2)$. Thus,
\begin{eqnarray}
	k_+^0=\frac{1}{2}\Big(E_{{q}+\frac{k_-}{2}}+E_{{q}-\frac{k_-}{2}}\Big)&\approx&(M^2+|\bm q|^2)^{1/2}+\frac{\left(M^2+|\bm q|^2\right)|\bm k_-|^2-(\bm q \cdot k_-)^2}{8 \left(M^2+|\bm q|^2\right)^{3/2}}+O\left(|\bm k_-|^3\right)\nonumber\\
    &=& E_q +\Delta E_q,
\end{eqnarray} 
where 
$\Delta E_q=\frac{E_q^2|\bm k_-|^2-(\bm q \cdot \bm k_{-})^2}{8E_q^3}$. Furthermore,
\begin{eqnarray}\nonumber
\delta(q^0-k^0_+)&=&\delta\bigg(q_0-E_q-\Delta E_q\bigg)
\\\nonumber
&=&2(E_q+\Delta E_q)\delta(q^2-M^2-2E_q\Delta E_q)
\\
&\approx&2E_q\delta(q^2-M^2)+2\Delta E_q\delta(q^2-M^2)-4E_q^2\Delta E_q\delta'(q^2-M^2), \label{eq:deltaexpansion}
\end{eqnarray}
and
\begin{eqnarray} \nonumber
&&\Big[\Big(|\bm q|^2+\frac{|\bm k_-|^2}{4}\Big)^2-(\bm q\cdot \bm k_{-})^2+2M^2\Big(|\bm q|^2+\frac{|\bm k_-|^2}{4}\Big)+M^4\Big]^{-1/4}
\\\nonumber
 &&\approx\frac{1}{(|\bm q|^2+M^2)^{1/2}}+\frac{(-M^2-|\bm q|^2+2|\bm q|^2 \cos[\theta]^2)|\bm k_-|^2}{8(|\bm q|^2+M^2)^{5/2}}
\\
&&=\frac{1}{E_q}\Big[1-\frac{\Delta E_q}{E_q}+\frac{(\bm q\cdot\bm k_-)^2}{8E_q^4}\Big].
 ~~~~\label{eq:predenofac}	
 \end{eqnarray}
We shall consider only the presence of diagonal spin components by postulating 
\begin{eqnarray}
\Big\langle a^{\dagger}\Big(\lambda', \bm q-\frac{\bm k_-}{2}\Big)a\Big(\lambda, \bm q+\frac{\bm k_-}{2}\Big)\Big\rangle\propto \delta_{\lambda\lambda'} 
\end{eqnarray} 
and similarly for anti-particles.
Now, the polarization-related structure appearing in the Wigner function can be approximated as 
\begin{eqnarray}	\epsilon_{\mu}\left(\lambda,q+\frac{k_-}{2}\right)\epsilon^{*}_{\nu}\left(\lambda,q-\frac{k_-}{2}\right) 
	\approx \Pi_{\mu\nu}^{(0)}(\lambda,q) + \frac{k_-^{\alpha}}{2} \Pi_{\mu\nu\alpha}^{(1)}(\lambda,q)+\frac{1}{2!}\frac{k_-^{\alpha}}{2}\frac{k_-^{\beta}}{2}\Pi^{(2)}_{\mu\nu\alpha\beta}(\lambda,q)
	\, \label{epsmn1}
\end{eqnarray}
up to $\mathcal{O}(k^2_-)$,
where 
\begin{eqnarray}
	\Pi_{\mu\nu}^{(0)}(\lambda,q)\equiv\epsilon_{\mu}(\lambda,q)\epsilon^{*}_{\nu}(\lambda,q)\,, \label{Pi0}
\end{eqnarray}
\begin{eqnarray}
\Pi_{\mu\nu\alpha}^{(1)}(\lambda,q)\equiv \big(\partial_{q^{\alpha}}\epsilon_{\mu}(\lambda,q)\big)\epsilon^{*}_{\nu}(\lambda,q)-\epsilon_{\mu}(\lambda,q)\big(\partial_{q^{\alpha}}\epsilon^{*}_{\nu}(\lambda,q)\big)\,,  \label{Pi1}
\end{eqnarray}
and
\begin{eqnarray}
\Pi_{\mu\nu\alpha\beta}^{(2)}(\lambda,q)&\equiv& \big(\partial_{q^{\alpha}}\partial_{q^{\beta}}\epsilon_{\mu}(\lambda,q)\big)\epsilon^{*}_{\nu}(\lambda,q)+\epsilon_{\mu}(\lambda,q)\big(\partial_{q^{\alpha}}\partial_{q^{\beta}}\epsilon^{*}_{\nu}(\lambda,q)\big)\nonumber\\&&-\big(\big(\partial_{q^{\alpha}}\epsilon_{\mu}(\lambda,q)\big)\big(\partial_{q^{\beta}}\epsilon^{*}_{\nu}(\lambda,q)\big)+\big(\partial_{q^{\beta}}\epsilon_{\mu}(\lambda,q)\big)\big(\partial_{q^{\alpha}}\epsilon^{*}_{\nu}(\lambda,q)\big)\big). \label{Pi2}
\end{eqnarray}

For simplicity, we will hereafter neglect the contribution for anti-particles. That is, we do not show the similar derivation for Wigner functions associated with $\langle b^{\dagger}b\rangle$. For the $\phi$ meson, which is its own anti-particle, such a contribution 
automatically vanishes. The Wigner function for vector mesons accordingly takes the form
\begin{eqnarray}
	W^{<\mu\nu}(q,X)&=&\sum_{\lambda=\pm 1,0}W^{<\mu\nu}(\lambda, q,X),
\end{eqnarray}
where
\begin{eqnarray} \nonumber
	W^{<\mu\nu}(\lambda, q,X)
	&=&\pi\int\frac{d^3k_{-}}{(2\pi)^3}{e^{-ik_-\cdot X}}\frac{1}{E_q}\Big[1-\frac{\Delta E_q}{E_q}+\frac{(\bm q\cdot\bm k_-)^2}{8E_q^4}\Big]
	\\\nonumber	&&\times\bigg[\left(\Pi^{\mu\nu}_{(0)}(\lambda,q) + \frac{k_{-\alpha}}{2} \Pi^{\mu\nu\alpha}_{(1)}(\lambda,q)+\frac{1}{2!}\frac{k_{-\alpha}}{2}\frac{k_{-\beta}}{2}\Pi_{(2)}^{\mu\nu\alpha\beta}(\lambda,q)\right) \nonumber 
    \\\nonumber  
	&&\times\left(2(E_q+\Delta E_q)\delta(q^2-M^2)-4E_q^2\Delta E_q\delta'(q^2-M^2)\right)
 \\
 &&\times\Big\langle a^{\dagger}\Big(\lambda, \bm q-\frac{\bm k_-}{2}\Big)a\Big(\lambda, \bm q+\frac{\bm k_-}{2}\Big)\Big\rangle
 \bigg].\label{eq:Wigner_V2}
\end{eqnarray}
For our purpose, we are interested in the on-shell Wigner function,
\begin{eqnarray}
W^{<\mu\nu}(\lambda,{\bm q},X)&=&\int \frac{{\rm d}q_0}{(2\pi)}W^{<\mu\nu}(\lambda,q,X), 
\end{eqnarray}
from which we introduce
\begin{eqnarray}
	W^{<\mu\nu}(\bm q,X)&=&\sum_{\lambda=\pm 1,0}W^{<\mu\nu}(\lambda, \bm q,X),
\end{eqnarray}
for its polarization averaged component. 
Keeping all the terms up to $\mathcal{O}(\bm k_-^2)$ and carrying out the integration over $\bm k_-$, one arrives at 
\begin{eqnarray}
W^{<\mu\nu}(\lambda,{\bm q},X)\sim
	\frac{1}{2 E_q}\left(\Pi^{\mu\nu}_{(0)}(\lambda,q) + \frac{i\hbar}{2}\Pi^{\mu\nu\alpha}_{(1)}\partial_{\alpha}(\lambda,q)-\frac{\hbar^2}{8}\Pi_{(2)}^{\mu\nu\alpha\beta}(\lambda,q)\partial_{\alpha}\partial_{\beta}\right){\tilde{f}}_{\lambda}(\bm q,X), 
\label{eq:Wigner_V3}
\end{eqnarray}
where we have introduced the polarization dependent distribution functions,
\begin{eqnarray}
	{\tilde{f}}_{\lambda}(\bm q,X)=\bigg[1+\frac{\hbar^2}{8 M^2}\left(\nabla^2-\frac{(q\cdot\nabla)^2}{E^2_q}\left(1+\frac{M^2}{E^2_q}\right)\right)\bigg]{f}_{\lambda}(\bm q,X) \label{eq:pddf}
\end{eqnarray} 
stemming from the expectation values of number operators
\begin{eqnarray}
	f_{\lambda}(\bm q,X)=\int\frac{{\rm d}^3k_-}{(2\pi)^3} {\rm e}^{-{\rm i}k_-\cdot X}\Big\langle a^{\dagger}\Big(\lambda,\bm q-\frac{\bm k_-}{2}\Big)a\Big(\lambda,\bm q+\frac{\bm k_-}{2}\Big)\Big\rangle.
\end{eqnarray} 
The custom of redefining distribution functions within the $\hbar$ expansion is commonly adopted in the Wigner-function approach for constructing quantum kinetic theories \cite{Hidaka:2016yjf,Hattori:2020guh}.

Given our expression of the Wigner function, one may further utilize the Kadanoff-Baym equation in the real-time formalism to derive the quantum kinetic equations for tracking the phase-space evolution of $\tilde{f}_{\lambda}(\bm q,X)$~\cite{Blaizot:2001nr,Hidaka:2022dmn}. However, such a derivation would be complicated, with a realistic collision term depending on the details of the interaction. In practice, it is even technically challenging to numerically solve a classical kinetic equation for spin-averaged vector mesons. Instead of seeking a precise value of $\rho_{00}$, the purpose of this work is to estimate the order of magnitude of the effect that the second order gradients have on $\rho_{00}$. It will hence here be sufficient to adopt the polarization-averaged distribution function obtained from classical transport equations to calculate $\rho_{00}$ from the Wigner function up to second order gradients as will be discussed in the subsequent sections.

\section{Spin density matrix}
\label{sec:SPD}
Diagonal elements of the spin density matrix are defined by the formula~\cite{Wagner:2022gza}
\begin{equation}
\rho_{\lambda\lambda}(q)=\frac{\int d\Sigma_{X}\cdot q \,\epsilon_{\mu}(\lambda,q)\epsilon^{*}_{\nu}(\lambda,q)W^{<\mu\nu}(\bm q,X)}{\int d\Sigma_{X}\cdot q\,\sum_{\lambda'=\pm 1,0}\,\epsilon_{\mu}(\lambda',q)\epsilon^{*}_{\nu}(\lambda',q)W^{<\mu\nu}(\bm q,X)}, \label{diag_spindensity}
\end{equation}
where $d\Sigma_{X\mu}$ denotes a freeze-out hyper-surface. 
The 00-component of the spin density matrix  
can further be expressed as
\begin{equation}
\rho_{00}(q)=\frac{\int d\Sigma_{X}\cdot q \,\Big[\Pi_{\mu\nu}^{(0)}(0,q)W^{<\mu\nu}(\bm q,X)\Big]}{\int d\Sigma_{X}\cdot q\,\Big[\Pi_{\mu\nu}^{(0)}(0,q)+\Pi_{\mu\nu}^{(0)}(-1,q)+\Pi_{\mu\nu}^{(0)}(+1,q)\Big]W^{<\mu\nu}(\bm q,X)}. \label{00_spindensity2}
\end{equation}
We can further decompose the Wigner function into symmetric and anti-symmetric components,
\begin{eqnarray}
\tilde{W}^{<\mu\nu}=\tilde{W}_{S}^{<\mu\nu}+i\tilde{W}_A^{<\mu\nu},
\end{eqnarray}
where 
\begin{eqnarray}
\tilde{W}_{S}^{<\mu\nu}\equiv\frac{1}{2}(\tilde{W}^{<\mu\nu}+\tilde{W}^{<\nu\mu}),\quad \tilde{W}_{A}^{<\mu\nu}\equiv\frac{-i}{2}(\tilde{W}^{<\mu\nu}-\tilde{W}^{<\nu\mu}).
\end{eqnarray}
Here, $\tilde{W}_{S}^{<\mu\nu}$ and $\tilde{W}_A^{<\mu\nu}$ are both real functions since $\tilde{W}^{<\mu\nu}$ by definition is Hermitian. 

Thus, $\rho_{00}$ can then be written as
\begin{equation}
\rho_{00}(q)=\frac{\int d\Sigma_{X}\cdot q \,\big(\epsilon_{(\mu}(0,q)\epsilon^{*}_{\nu)}(0,q)W_S^{<\mu\nu}(\bm q,X)+i\epsilon_{[\mu}(0,q)\epsilon^{*}_{\nu]}(0,q)W_A^{<\mu\nu}(\bm q,X)\big)}{\int d\Sigma_{X}\cdot q\,\sum_{\lambda=\pm 1,0}\,\big(\epsilon_{(\mu}(\lambda,q)\epsilon^{*}_{\nu)}(\lambda,q)W_S^{<\mu\nu}(\bm q,X)+i\epsilon_{[\mu}(\lambda,q)\epsilon^{*}_{\nu]}(\lambda,q)W_A^{<\mu\nu}(\bm q,X)\big)}.
\end{equation}
We may next re-express the distribution functions for $\lambda=\pm 1$ in terms of $\tilde{f}_{\pm 1}=f_V\pm f_A/2$, where $f_A=\tilde{f}_{1}-\tilde{f}_{-1}\neq 0$ characterizes a non-zero spin polarization of vector mesons. 
When averaging over the spin polarization, we may set $\tilde{f}_{\lambda}=f_V$ with $f_A=0$. In principle, as briefly explained in the previous section, we should work out the quantum kinetic equations for each $\tilde{f}_{\lambda}$ with $\hbar$ corrections in the presence of collisions. 
$\tilde{W}^{\mu\nu}(\lambda,q,X)$ may furthermore  possibly involve $\hbar$ corrections pertinent to interactions. However, for simplicity, we here only consider the quantum correction induced by the spacetime-derivative terms in Eq.~(\ref{eq:Wigner_V3}) obtained in the collisionless limit and incorporate collisional effects in $\tilde{f}_{\lambda}(q,X)$ only from polarization-averaged transport theory. Consequently, we will simply consider the case of $\tilde{f}_{\lambda}=f_V$ and vanishing $W^{\mu\nu}_{A}$ at $\mathcal{O}(\hbar^0)$.

Note that the essential source of spin alignment considered in this work is led by $\partial_{\alpha}f_V$ and $\partial_{\alpha}\partial_{\beta}f_V$. Since  $\rho_{00}$ is a normalized quantity, an overall factor of the Wigner function should not  matter. To obtain a more concrete expression of $\rho_{00}$, we shall further compute the coefficients associated with the momentum derivatives on the polarization vectors explicitly. 
%
%
\subsection{Decomposition of $\Pi^{\mu\nu}_{(0)}(\lambda,q)$, $\Pi^{\mu\nu\alpha}_{(1)}(\lambda,q)$ and $\Pi^{\mu\nu\alpha\beta}_{(2)}(\lambda,q)$ in symmetric and antisymmetric parts}
We first decompose $\Pi^{\mu\nu}_{(0)}$ in symmetric and antisymmetric components  such that $\Pi^{\mu\nu}_{(0)}(\lambda,q)=\frac{1}{2}(\Pi^{(\mu\nu)}_{(0)}(\lambda,q)+\Pi^{[\mu\nu]}_{(0)}(\lambda,q))$ where 
\begin{eqnarray}
	\Pi^{(\mu\nu)}_{(0)}(\lambda,q)\equiv\epsilon^{(\mu}(\lambda,q)\epsilon^{*\nu)}(\lambda,q),
\end{eqnarray}
and
\begin{eqnarray}
	\Pi^{[\mu\nu]}_{(0)}(\lambda,q)\equiv\epsilon^{[\mu}(\lambda,q)\epsilon^{*\nu]}(\lambda,q).
\end{eqnarray}
Similarly, $\Pi_{\mu\nu\alpha}^{(1)}(\lambda,q)$ and $\Pi_{\mu\nu\alpha\beta}^{(2)}(\lambda,q)$ can be decomposed in symmetric and antisymmetric parts as 
\begin{eqnarray}
\Pi^{[\mu\nu]\alpha}_{(1)}(\lambda,q)=\big(\partial^{\alpha}_{q}\epsilon^{[\mu}(\lambda,q)\big)\epsilon^{*\nu]}(\lambda,q)+{\rm c.c.}, \label{Pimnaantisym}
\end{eqnarray}
\begin{eqnarray}
\Pi^{(\mu\nu)\alpha}_{(1)}(\lambda,q)=\big(\partial^{\alpha}_{q}\epsilon^{(\mu}(\lambda,q)\big)\epsilon^{*\nu)}(\lambda,q)-{\rm c.c.}, \label{Pimnasym}
\end{eqnarray}
\begin{eqnarray}
\Pi^{[\mu\nu]\alpha\beta}_{(2)}(\lambda,q)\equiv \Bigg(\left(\partial_{q}^{{\alpha}}\partial_{q}^{{\beta}}\epsilon^{[\mu}(\lambda,q)\right)\epsilon^{*\nu]}(\lambda,q)-\big(\big(\partial_{q}^{{\alpha}}\epsilon^{[\mu}(\lambda,q)\big)\big(\partial_{q}^{{\beta}}\epsilon^{*\nu]}(\lambda,q)\big)\big)\Bigg)-{\rm c.c.},  \label{Pimnabantisym}
\end{eqnarray}
\begin{eqnarray}
\Pi^{(\mu\nu)\alpha\beta}_{(2)}(\lambda,q)\equiv \Bigg(\left(\partial_{q}^{{\alpha}}\partial_{q}^{{\beta}}\epsilon^{(\mu}(\lambda,q)\right)\epsilon^{*\nu)}(\lambda,q)-\big(\big(\partial_{q}^{{\alpha}}\epsilon^{(\mu}(\lambda,q)\big)\big(\partial_{q}^{{\beta}}\epsilon^{*\nu)}(\lambda,q)\big)\big)\Bigg)+{\rm c.c.}, \label{Pimnabsym}
\end{eqnarray}
where c.c. represents the complex conjugate. 
We choose the $y$ direction as the spin quantization axis, in which case the corresponding spin state vectors read~\cite{Sheng:2022ffb}
\begin{eqnarray}
 {\bm \epsilon_0}&=&\Big(0,1,0\Big),~~~~~~~~~~~~~~~~~~~~~~ \label{e0}  \\
 {\bm \epsilon_{+1}}&=&-\frac{1}{\sqrt{2}}\Big(i,0,1\Big), ~~~~~~~~~~~~~~~~\label{e+1}\\ 
  {\bm \epsilon_{-1}}&=&\frac{1}{\sqrt{2}}\Big(-i,0,1\Big), ~~~~~~~~~~~~~~~\label{e-1}
\end{eqnarray}
which satisfy the relation 
\begin{equation}
\sum_{\lambda=\pm 1,0}{\bm\epsilon^m_{\lambda}\bm\epsilon^{* n}_{\lambda}}=\delta^{mn}. \label{eps_r2}
\end{equation}

Using Eqs.~(\ref{e0}-\ref{e-1}), one can show that the polarization tensor $\epsilon^{\mu}$ (see Eq.(\ref{pol_vec1})) satisfies
\begin{eqnarray}
\epsilon^{\mu}(0,q)=\epsilon^{*\mu}(0,q), \label{eq:epsprop1}\\
 \epsilon^{\mu}(1,q)=-\epsilon^{*\mu}(-1,q).   \label{eq:epsprop2}
\end{eqnarray}

Based on the above properties, we have
\begin{eqnarray}
\Pi^{(\mu\nu)}_{(0)}(1,q)=\Pi^{(\mu\nu)}_{(0)}(-1,q),\quad \label{eq:pmnsrel}\\
\Pi^{[\mu\nu]}_{(0)}(1,q)=-\Pi^{[\mu\nu]}_{(0)}(-1,q), \label{eq:pmnasrel}
\end{eqnarray}
and
\begin{eqnarray}
\Pi^{(\mu\nu)\alpha}_{(1)}(1,q)=\big(\partial^{\alpha}_{q}\epsilon^{(\mu}(1,q)\big)\epsilon^{*\nu)}(1,q)-{\rm c.c.}=-\Pi^{(\mu\nu)\alpha}_{(1)}(-1,q), \quad\label{eq:pmnalpsrel}\\
\Pi^{[\mu\nu]\alpha}_{(1)}(1,q)=\big(\partial^{\alpha}_{q}\epsilon^{[\mu}(1,q)\big)\epsilon^{*\nu]}(1,q)+{\rm c.c.}=\Pi^{[\mu\nu]\alpha}_{(1)}(-1,q),\quad\label{eq:pmnalpasrel}
\end{eqnarray}
\begin{eqnarray}
\Pi^{(\mu\nu)\alpha\beta}_{(2)}(1,q)&\equiv& \Bigg(\left(\partial_{q}^{{\alpha}}\partial_{q}^{{\beta}}\epsilon^{(\mu}(1,q)\right)\epsilon^{*\nu)}(1,q)-\big(\big(\partial_{q}^{{\alpha}}\epsilon^{(\mu}(1,q)\big)\big(\partial_{q}^{{\beta}}\epsilon^{*\nu)}(1,q)\big)\big)\Bigg)+{\rm c.c.} \nonumber\\&=& \Pi^{(\mu\nu)\alpha\beta}_{(2)}(-1,q), \quad \label{Pimnabsymrel}\\
\Pi^{[\mu\nu]\alpha\beta}_{(2)}(1,q)&\equiv& \Bigg(\left(\partial_{q}^{{\alpha}}\partial_{q}^{{\beta}}\epsilon^{[\mu}(1,q)\right)\epsilon^{*\nu]}(1,q)-\big(\big(\partial_{q}^{{\alpha}}\epsilon^{[\mu}(1,q)\big)\big(\partial_{q}^{{\beta}}\epsilon^{*\nu]}(1,q)\big)\big)\Bigg)-{\rm c.c.}\nonumber\\&=& -\Pi^{[\mu\nu]\alpha\beta}_{(2)}(-1,q). \quad  \label{Pimnabantisymrel}
\end{eqnarray} 

Moreover, using Eq.~(\ref{eq:epsprop1}), we can show that  
\begin{eqnarray}
  \Pi^{[\mu\nu]}_{(0)}(0,q)=0, \quad \label{eq:pmnas0q}\\
    \Pi^{(\mu\nu)\alpha}_{(1)}(0,q)=0, \quad \label{eq:pmnsa0q}\\
\Pi^{[\mu\nu]\alpha\beta}_{(2)}(0,q)=0. \quad \label{eq:pmnasab0q}
\end{eqnarray}

 Making use of Eqs.~(\ref{eq:pmnsrel}-\ref{eq:pmnasab0q}) for Eq.~(\ref{eq:Wigner_V3}) and substituting $\tilde{f}_{0,\pm1}=f_V$,  we obtain 
\begin{eqnarray}
W^{<\mu\nu}(q,X)&=&\frac{1}{2 E_q}\bigg[\frac{1}{2}\Pi^{(\mu\nu)}_{(0)}(q)+\frac{i\hbar}{4}\big(\Pi^{[\mu\nu]\alpha}_{(1)}(1,q)+\Pi^{[\mu\nu]\alpha}_{(1)}(-1,q)+\Pi^{[\mu\nu]\alpha}_{(1)}(0,q)\big)\partial_{\alpha}\nonumber\\
&&-\frac{\hbar^2}{16}\big(\Pi^{(\mu\nu)\alpha\beta}_{(2)}(1,q)+\Pi^{(\mu\nu)\alpha\beta}_{(2)}(-1,q)+\Pi^{(\mu\nu)\alpha\beta}_{(2)}(0,q)\big)\partial_{\alpha}\partial_{\beta}\bigg]f_V(q,X), \quad \label{tot_wig_fn}
\end{eqnarray}
where $\Pi^{(\mu\nu)}_{(0)}(q)=\sum_{\lambda=\pm 1,0}\Pi^{(\mu\nu)}_{(0)}(\lambda,q)=2\left(\frac{q^{\mu}q^{\nu}}{M^2}-\eta^{\mu\nu}\right)$.

Eventually, $\rho_{00}$ reads
\begin{eqnarray}
\rho_{00}(q)
&=&\frac{\int d\Sigma_{X}\cdot q \,\bigg[1-\frac{\hbar^2}{32}\Pi_{(\mu\nu)}^{(0)}(0,q)\Pi^{(\mu\nu)\alpha\beta}_{(2)}(q)\partial_{\alpha}\partial_{\beta}\bigg]f_V(q,X)}{\int d\Sigma_{X}\cdot q\,\bigg[3-\frac{\hbar^2}{32}\Pi_{(\mu\nu)}^{(0)}(q)\Pi^{(\mu\nu)\alpha\beta}_{(2)}(q)\partial_{\alpha}\partial_{\beta}\bigg]f_V(q,X)}, \label{eq:rho00q}
\end{eqnarray}
where $\Pi_{(\mu\nu)}^{(0)}(0,q)\Pi^{(\mu\nu)\alpha\beta}_{(2)}(q)\partial_{\alpha}\partial_{\beta}$ and $\Pi_{(\mu\nu)}^{(0)}(q)\Pi^{(\mu\nu)\alpha\beta}_{(2)}(q)\partial_{\alpha}\partial_{\beta}$, can be written as 
\begin{eqnarray}
\Pi_{(\mu\nu)}^{(0)}(0,q)\Pi^{(\mu\nu)\alpha\beta}_{(2)}(q)\partial_{\alpha}\partial_{\beta}&=&\bigg[\frac{2(q^y)^2}{M^2}A_1-\frac{4(q^y)^2E_q}{M^2}B_2\bigg]\partial_{\alpha}\partial^{\alpha}
\nonumber\\&&
+\bigg(4(B_3+C_3)E_q-4(A_2+C_2)\bigg)\bigg[\frac{q^y}{M}\left(\partial^{y}+\frac{q^y(q^i\partial^{i})}{M(E_q+M)}\right)\bigg]q^{\alpha}\partial_{\alpha}
\nonumber\\&&
+\bigg[\frac{4(q^{y} E_q)^2}{M^2}E_3-\frac{4\bm (q^y)^2}{M}\frac{E_q}{M}D_2\bigg](q^{\alpha}\partial_{\alpha})^2
\nonumber\\&&
+4A_3\bigg[(\partial^{y})^2+\frac{2q^y}{M(E_q+M)}(q^i\partial^{i})\partial^{y}+\frac{(q^y)^2}{M^2(E_q+M)^2}(q^i \partial^i)^2\bigg]~~~~\label{eq:rho_fac_num}
\end{eqnarray}
and
\begin{eqnarray}
\Pi_{(\mu\nu)}^{(0)}(q)\Pi^{(\mu\nu)\alpha\beta}_{(2)}(q)\partial_{\alpha}\partial_{\beta}&=&2\bigg[\left(\left(\frac{E_q^2}{M^2}-1\right)A_1-\frac{2|\bm q|^2}{M^2}E_qB_2 \right)\partial_{\alpha}\partial^{\alpha}+2A_3(\partial^{i}\partial^{i})+\frac{2}{M^2}A_3(q^{i}\partial^{i})^2
\nonumber\\&&+\frac{E_q}{M^2}\left(2E_q(B_3+C_3)-2(A_2+C_2)\right)(q^{i}\partial^{i})(q^{\beta}\partial_{\beta})
\nonumber\\&&+E_q\left(\frac{|\bm q|^2}{M^2}\right)\left(2E_qE_3-2D_2\right)(q^{\beta}\partial_{\beta})^2\bigg]. \label{eq:rho_fac_deno}
\end{eqnarray}
The coefficients $A_{1-3}$, $B_{2-3}$, $C_{2-3}$, $D_2$ and $E_3$ are listed in Appendix~\ref{app:components_Pi2}. We shall in the next Section examine the large-momentum and low-momentum limits, in which $\rho_{00}$ assumes a 
greatly simplified form.
\subsection{Large-momentum limit}
At large momentum, $|\bm q|>> M$ and we have $E_q\sim |\bm q|$, which leads to
\begin{eqnarray}
\Pi_{(\mu\nu)}^{(0)}(0,q)\Pi^{(\mu\nu)\alpha\beta}_{(2)}(q)\partial_{\alpha}\partial_{\beta}&\approx&
-\frac{4(q^y)^2}{M^4|\bm q|^2}\bigg[3|\bm q|^2(\partial_{t})^2-|\bm q|^2(\partial^{i})^2-4\left(q^i\partial^{i}\right)q^{\alpha}\partial_{\alpha}
\bigg],\label{eq:53}
\end{eqnarray}
and
\begin{eqnarray}
\Pi_{(\mu\nu)}^{(0)}(q)\Pi^{(\mu\nu)\alpha\beta}_{(2)}(q)\partial_{\alpha}\partial_{\beta}&\approx&
-\left(\frac{4}{M^4}\right)\bigg[3|\bm q|^2(\partial_{t})^2-|\bm q|^2(\partial^{i})^2-4\left(q^i\partial^{i}\right)q^{\alpha}\partial_{\alpha}
\bigg]. \label{eq:54}
\end{eqnarray}
Using Eqs.~(\ref{eq:53}-\ref{eq:54}), we can simplify Eq.~(\ref{eq:rho00q}) to 
\begin{eqnarray}
\rho_{00}(q)
&=&\frac{\int d\Sigma_{X}\cdot q \,\Big[1+\frac{\hbar^2}{8}\frac{(q^y)^2}{M^4}{\slashed{\Box}}\Big]f_V(q,X)}{\int d\Sigma_{X}\cdot q\,\Big[3+\frac{\hbar^2}{8}\left(\frac{|\bm q|^2}{M^4}\right){\slashed{\Box}}\Big]f_V(q,X)}, \label{eq:rho001}
\end{eqnarray}
where
\begin{eqnarray}
{\slashed{\Box}}=\bigg(3(\partial_{t})^2-(\partial^{i})^2-\frac{4}{|\bm q|^2}\left(q^i\partial^{i}\right)q^{\alpha}\partial_{\alpha}
\bigg).
\end{eqnarray}
The above result shows that $\rho_{00}-1/3$ at large momentum could be substantially enhanced when $\mathcal{O}\big(|\bm q|^2\partial^2f_V/(M^{4}f_V)\big)\sim \mathcal{O}(1)$.

\subsection{Small-momentum limit}
In the small momentum limit, $|\bm q|<< M$, we have $E_q\sim M$, giving
\begin{eqnarray}
\Pi_{(\mu\nu)}^{(0)}(0,q)\Pi^{(\mu\nu)\alpha\beta}_{(2)}(q)\partial_{\alpha}\partial_{\beta}&\approx&
4A_3(\partial^{y})^2=2(\partial^{y})^2/M^2,
\end{eqnarray}
and
\begin{eqnarray}
\Pi_{(\mu\nu)}^{(0)}(q)\Pi^{(\mu\nu)\alpha\beta}_{(2)}(q)\partial_{\alpha}\partial_{\beta}&\approx&4A_3(\partial^{i})^2
=2(\partial^{i})^2/M^2.
\end{eqnarray}
Thus, $\rho_{00}(q)$ for this case becomes
\begin{eqnarray}
\rho_{00}(q)
&=&\frac{\int d\Sigma_{X}\cdot q \,\bigg[1-\frac{\hbar^2}{4}\frac{(\partial^{y})^2}{M^2}\bigg]f_V(q,X)}{\int d\Sigma_{X}\cdot q\,\bigg[3-\frac{\hbar^2}{4}\frac{(\partial^{i})^2}{M^2}\bigg]f_V(q,X)}, \label{eq:rho0011}
\end{eqnarray}
which can be further approximated as 
\begin{eqnarray}
\rho_{00}(q)
&=&\frac{1}{3}-\frac{\hbar^2 }{12M^2}\frac{\int d\Sigma_{X}\cdot q \,(2(\partial^{y})^2-(\partial^{x})^2-(\partial^{z})^2)f_V(q,X)}{\int d\Sigma_{X}\cdot q \,f_V(q,X)},
\label{eq:rho002}
\end{eqnarray}
when $\mathcal{O}\big(\partial^2f_V/(M^2f_V)\big)\ll 1$.
%
%
\section{Spin alignment observables within a thermal model with single freeze out}
\label{sec:thm}
In order to estimate $\rho_{00}$, we assume the vector mesons to follow the J{\"u}ttner distribution, ($f_V(q,X)=\exp[-q^{\mu}\beta_{\mu}(x)-\xi(x)]$ where, $\beta^\mu=u^{\mu}/T$ and $\xi=\mu/T$ are ratios of the four fluid velocity and the chemical potential to the temperature) and use  a thermal model with single-freeze-out ~\cite{Broniowski:2001we} which has been used in the past to describe various features of soft hadron production (particle yields, transverse-momentum spectra, elliptic flow, HBT radii) for Au+Au collisions at the top RHIC energies ($\sqrt{s_{NN}}=130$ GeV)~\cite{Cleymans:1992zc,Braun-Munzinger:2001hwo,Florkowski:2001fp,Becattini:2005xt,Andronic:2017pug}.
In its standard formulation,  it uses two thermodynamic parameters (temperature $T$, baryon chemical potential $\mu_B$) and two geometric parameters (proper time $\tau_f$ and system size $r_{\rm max}$) which characterize the freeze-out hypersurface (defined through the conditions: $\tau^2_f = t^2 - x^2 - y^2 - z^2$ and $x^2 + y^2 \leq r^2_{\rm max}$) and hydrodynamic flow (which is assumed to have a Hubble-like form, $u^\mu = x^\mu/\tau$). The thermodynamic parameters $T$, $\mu_B$ are obtained by fitting the ratios of hadronic abundances to experimental data while the geometric ones ($\tau_f$ and $r_{\rm max}$) are determined by fits to experimental transverse-momentum spectra.
In this work, we use the extended thermal model with a single freeze out, in which phenomena such as elliptic flow are included by taking into account the elliptic deformations of both the emission region in the transverse plane and the transverse flow~\cite{Broniowski:2002wp} 
in terms of two new parameters $\epsilon$ and $\delta$. To include the elliptical asymmetry in the transverse plane, the transverse region is modeled by the parameterization 
\beq 
x &=& r_{\rm max} \sqrt{1-\epsilon} \cos\phi, \nn \\
y &=& r_{\rm max} \sqrt{1+\epsilon} \sin\phi,
\eeq 
where $\phi$ is the azimuthal angle, while $r_{\rm max}$ and $\epsilon$ are the model parameters. $\epsilon >0$ indicates that the system formed in the collision is elongated in the $y$ direction.

Flow asymmetry is included by parametrizing the flow velocity as
\beq
u^\mu&=&\frac{1}{N}\left(t,~x\sqrt{1+\delta},~ y\sqrt{1-\delta },~z\right), \label{eq:u}
\eeq
where the parameter $\delta$ characterizes the anisotropy of the  transverse flow. $\delta > 0$, indicates that there is more flow in the reaction plane (elliptic flow). 
The constant $N$ can be obtained from the normalization condition $u^{\mu}u_{\mu}=1$, yielding 
\beq
N={\sqrt{\tau ^2-\left(x^2-y^2\right)\delta}}\,,
\label{eq:N}
\eeq
where the proper time $\tau$ is given by 
\beq
\tau^2 = t^2 -x^2-y^2-z^2. 
\label{eq:tau}
\eeq
The adopted parameter values for $\epsilon$, $\delta$, $\tau_f$, and $r_{\rm max}$ are listed in Table~\ref{tab}. 
These values have been used in the past to describe the PHENIX data for the centrality classes 
$c$=0--15\%, $c$=15--30\% and $c$=30--60\% at beam energy $\sqrt{s_{NN}}=130$~GeV and freeze-out temperature $T_f=0.165$~GeV \cite{baran, Florkowski:2004du}.
\begin{table*}[ht!]
\centering
\begin{tabular}{ |p{3cm}||p{3cm}||p{3cm}||p{3cm}||p{3cm}|} 
   \hline
 c $\%$ & $\epsilon$ &   $\delta$ &   $\tau_f$~[fm] &   $r_{\rm max}$~[fm] \\
  \hline
   $0-15$ &   $0.055$ &   $0.12$ &   $7.666$ &   $6.540$ \\
  $15-30$ &   $0.097$ &   $0.26$ &   $6.258$ &   $5.417$ \\
  $30-60$ &   $0.137$ &   $0.37$ &   $4.266$ &   $3.779$ \\
 \hline
\end{tabular}
\caption{Values of the parameters previously used to describe the PHENIX data at $\sqrt{s_{NN}}=130$~GeV for different centrality classes~\cite{baran,Florkowski:2004du}. The freeze-out temperature used in the calculation is $T_f =$~165~MeV.}
\label{tab}
\end{table*}
 We furthermore assume that the freeze-out takes place at a constant value of the proper time, i.e., at $\tau = \tau_f$. In this case, the element of the freeze-out hypersurface, $d\Sigma_{X\lambda}$, can be expressed as 
\begin{eqnarray}
d\Sigma_{X\lambda } &=& n_{\lambda }\, dx dy\,  d\eta \label{sig}, 
\end{eqnarray}
where
\begin{eqnarray}
n^{\lambda } = \left(\sqrt{\tau^2_f+x^2+y^2}\cosh\eta,x,y,
\sqrt{\tau^2_f + x^2+y^2}\sinh\eta \right), \nonumber\\
\label{eq:n}
\end{eqnarray}
with \mbox{$n^\lambda n_\lambda = \tau^2_f$} and $\eta=\frac{1}{2} \ln \left[(t+z)/(t-z)\right]$ being the space-time rapidity.

The particle four momentum reads 
\begin{eqnarray}
q^{\mu}=\left( E_q,q_x,q_y,q_z \right) = \left( m_T\ch y_p, q_x ,q_y, m_T\sh y_p \right), \lab{ql}
\end{eqnarray}
where $m_T = \sqrt{m^2 + q_T^2}$ is the transverse mass, $q_T=\sqrt{q_x^2+q_y^2}$ and $m$ being the transverse momentum and mass of the particle, respectively. $y_p=\frac{1}{2}\ln \left[(E_q+q_z)/(E_q-q_z)\right]$ is the particle rapidity. 
Using Eqs.~(\ref{eq:u}) and (\ref{ql}) we can write, 
\begin{eqnarray}
q^{\mu}\beta_{\mu}=R_1 \cosh(y_p-\eta)+R_2, \label{eq:bq}
\end{eqnarray}
where 
\begin{eqnarray}
R_1&=&\frac{m_T\left(\cosh(y_p) t-\sinh(y_p) z\right)}{T_{f}\sqrt{t^2-x^2-y^2-z^2-(x^2-y^2)\delta}}, \label{eq:R1}\\
R_2&=&-\frac{\left(x\sqrt{1+\delta}\, q_x+y\sqrt{1-\delta}\,q_y\right)}{T_{f}\sqrt{t^2-x^2-y^2-z^2-(x^2-y^2)\delta}}.\label{eq:R2}
\end{eqnarray}
Accordingly, we have 
\begin{eqnarray}
d \Sigma_{X}\cdot q&=& \left(G_1\cosh(y_p-\eta)+G_2\right) dx\,dy\,d\eta,  \label{eq:dsp}
\end{eqnarray}
where
\begin{eqnarray}
G_1&=&m_T\sqrt{\tau_f^2+x^2+y^2}, \label{eq:G1}\\
G_2&=&-\left(x q_x+y q_y\right).\label{eq:G2}
\end{eqnarray}
At the top RHIC energies we can neglect effects of the baryon number density and  calculate the contribution from temperature gradients using  the hydrodynamic equations within the perfect fluid approximation (see appendix A of Ref.~\cite{Florkowski:2019voj} for details).
In such a situation, using Eqs.~(\ref{eq:rho_fac_num}-\ref{eq:rho_fac_deno}),  we can easily carry out the freezeout integration in (\ref{eq:rho00q}) and plot $\delta\rho_{00}(q_x,q_y)=\rho_{00}(q_x,q_y)-\frac{1}{3}$ as a function of $q_x$ and $q_y$. 
The azimuthal angle dependence $\rho_{00}(\phi_p)$ (at fixed particle rapidity $y_p=0$) can be obtained by 
taking the momentum average of $\rho_{00}(q)$ weighted by the momentum spectrum of the considered meson~\cite{Sheng:2022wsy,Sheng:2023urn},
\begin{eqnarray}
\langle \rho_{00}(\phi_q)\rangle&=& \frac{\int q_T dq_T \left[\rho_{00}(\mathbf{q})\cal{N}\right]}{\int dq_T  q_T \cal{N}}, 
\label{eq:rho_phi}
\end{eqnarray}
where ${\cal{N}}=\int d\Sigma_{X} \cdot q \,f_V(q,X)$. 
One can also look at $\rho_{00}$ as a function of $q_T$ at $y_p=0$. In this case, we perform the azimuthal-angle integrals in both the numerator and denominator, keeping the transverse momentum fixed, namely
\begin{eqnarray}
\langle \rho_{00}(q_T)\rangle&=&\frac{\int d\phi_p \left[\rho_{00}(\mathbf{q})\cal{N}\right]}{\int d\phi_p {\cal{N}}}.\label{eq:P_pT}
\end{eqnarray}
Finally, one can also obtain the rapidity dependence of $\rho_{00}$ from the formula 
\begin{eqnarray}
\langle \rho_{00}(y_p)\rangle&=&\frac{\int q_T dq_T\int d\phi_p \left[\rho_{00}(\mathbf{q})\cal{N}\right]}{\int q_T dq_T\int d\phi_p {\cal{N}}}.\label{eq:y_p}
\end{eqnarray}
\section{Results and discussions}
\label{sec:res}
In this section we present the numerical results for $\phi$ and $K^{*0}$ mesons, making use of the thermal model parameters 
of Table~\ref{tab}.
 
\begin{figure*}
      \centering
	{}\includegraphics[width=0.48\textwidth]{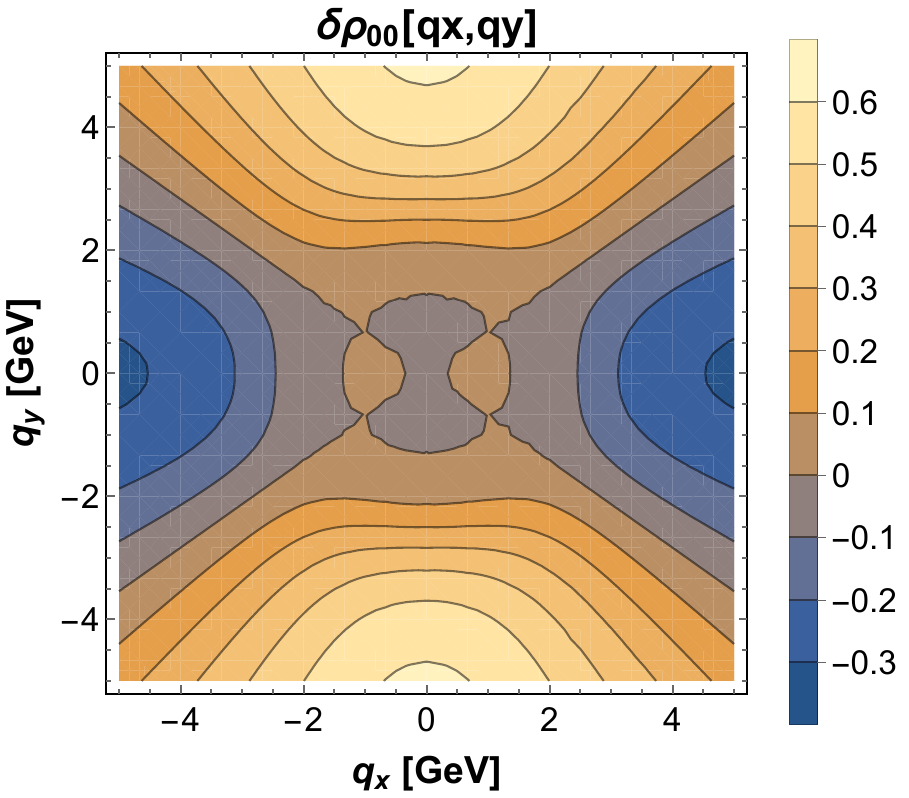}
	{}\includegraphics[width=0.48\textwidth]{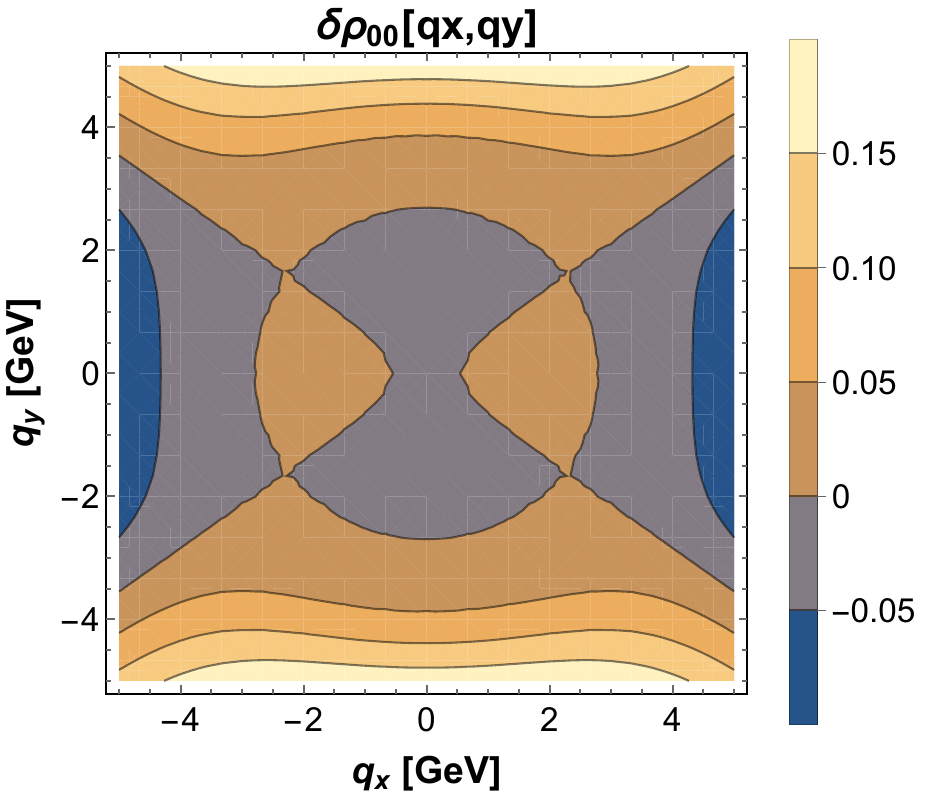}
        {}\includegraphics[width=0.48\textwidth]{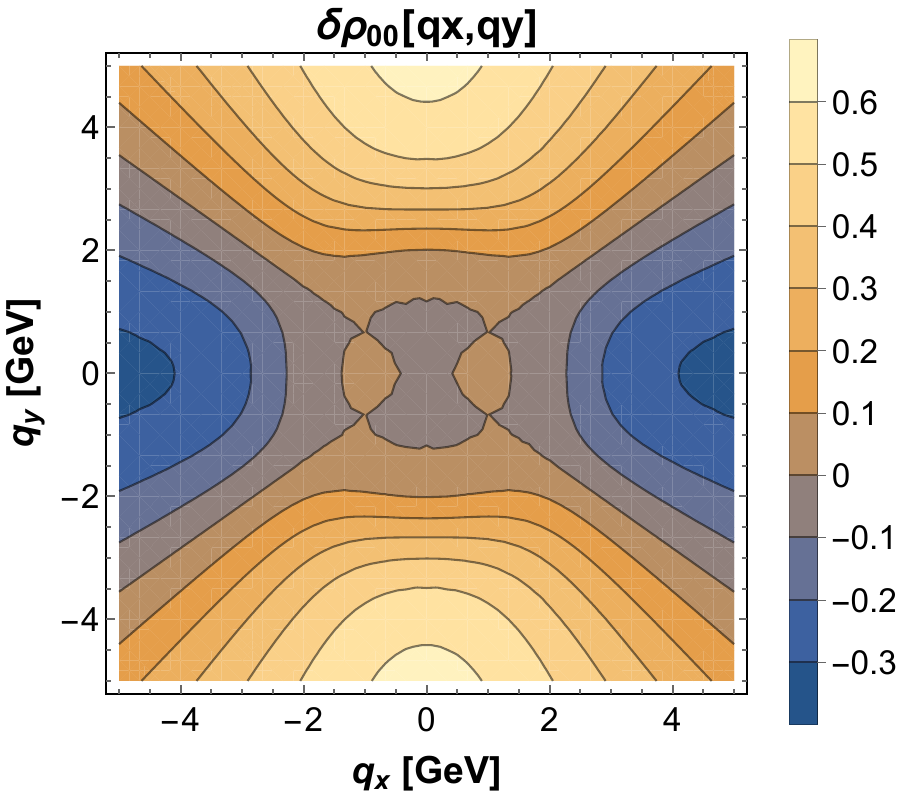}
        {}\includegraphics[width=0.48\textwidth]{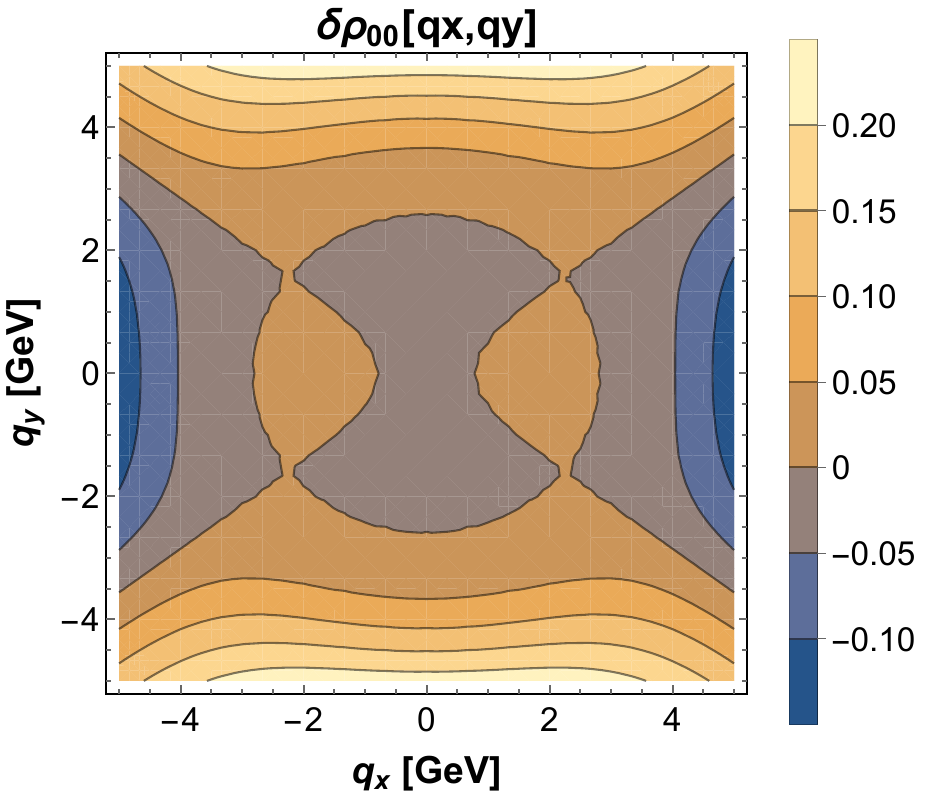}
        {}\includegraphics[width=0.48\textwidth]{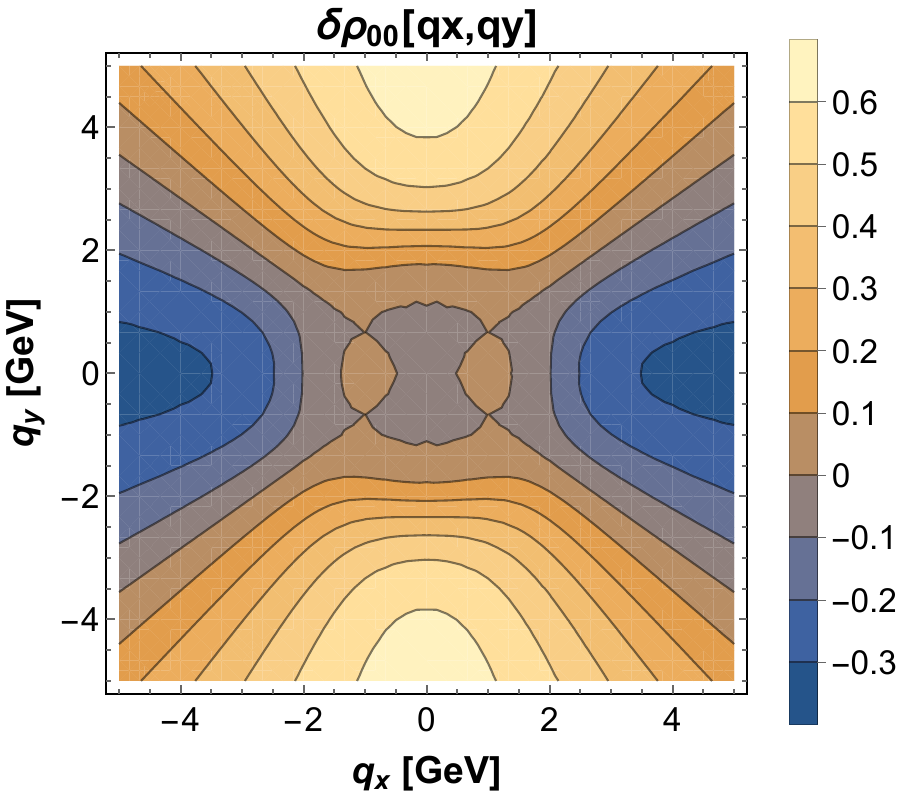}
        {}\includegraphics[width=0.48\textwidth]{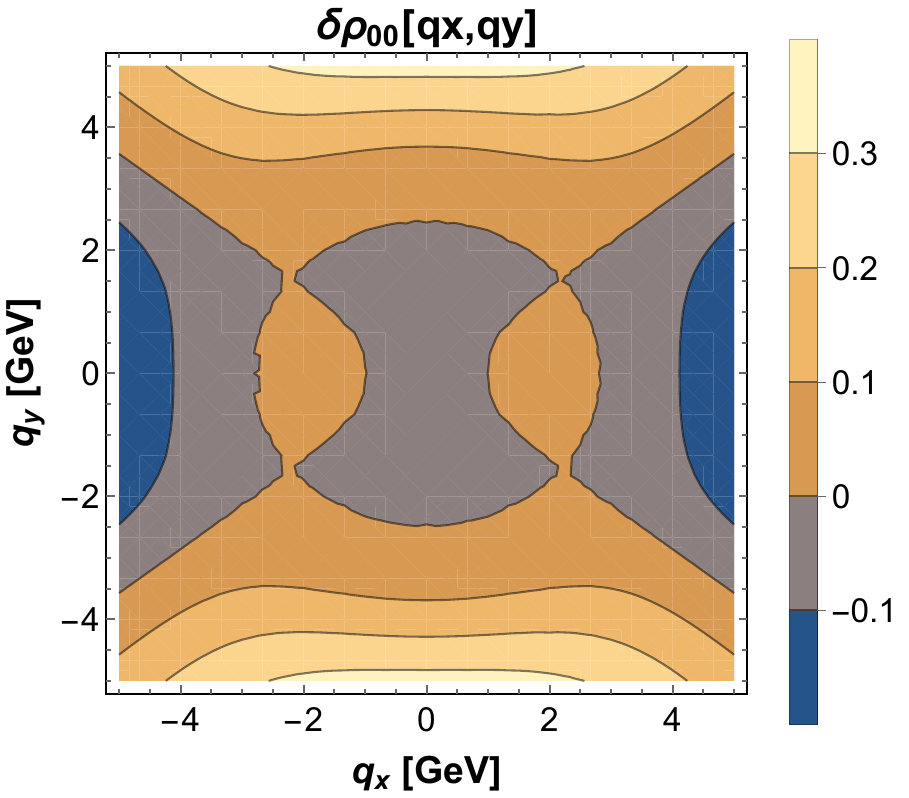}
		\caption{The contour plots of $\delta\rho_{00} = \rho_{00} - 1/3$ as a function of $q_x$ and $q_y$ for $K^{*0}$ (left panels)  and the $\phi$ meson (right panels) for the centrality classes $c$=0--15\%,  $c$=15--30\% and  $c$=30--60\% (top to bottom) with freeze-out temperature $T_f=165$ MeV and collision energy $\sqrt{s_{NN}}=130$ GeV at $y_p=0$.
		}
	 \label{fig:f1}
\end{figure*}

In FIG.~\ref{fig:f1}, we show the contour plot of $\delta\rho_{00}$ for $K^{*0}$ (left panel from top to bottom) and $\phi$ (right panel from top to bottom) as a function of $q_x$ and $q_y$ for the three centrality classes $c=0-15\%$, $c=15-30\%$ and $c=30-60\%$ at $y_p=0$. For $K^{*0}$ mesons  we  observe a significant quadrupole structure for $\delta\rho_{00}$ which changes sign as a function of $q_x$ and $q_y$. At smaller $q_x$, $q_y$ (in the domain  $q_x\in(-1.5,1.5)$ GeV and $q_y\in(-1.5,1.5)$ GeV) when $|q_y|>>|q_x|$ we have $\delta\rho_{00}<0$ while for $|q_y|<<|q_x|$ such that $0.5<|q_x|<1.5$ we have $\delta\rho_{00}>0$. 
On the other hand, at higher momentum ($|q_x|,|q_y| >1.5$ GeV) when $|q_x|>>|q_y|$, we see $\delta\rho_{00}<0$ while for $|q_y|>>|q_x|$, $\delta\rho_{00}>0$. A similar pattern for the $\phi$ meson with slightly different ranges of $q_x$ and $q_y$ is observed. We note that, albeit the complicated structures, the difference in the results for $K^{*0}$ and $\phi$ interestingly only stem from their mass difference.

\begin{figure*}
      \centering
        {}\includegraphics[width=0.45\textwidth]{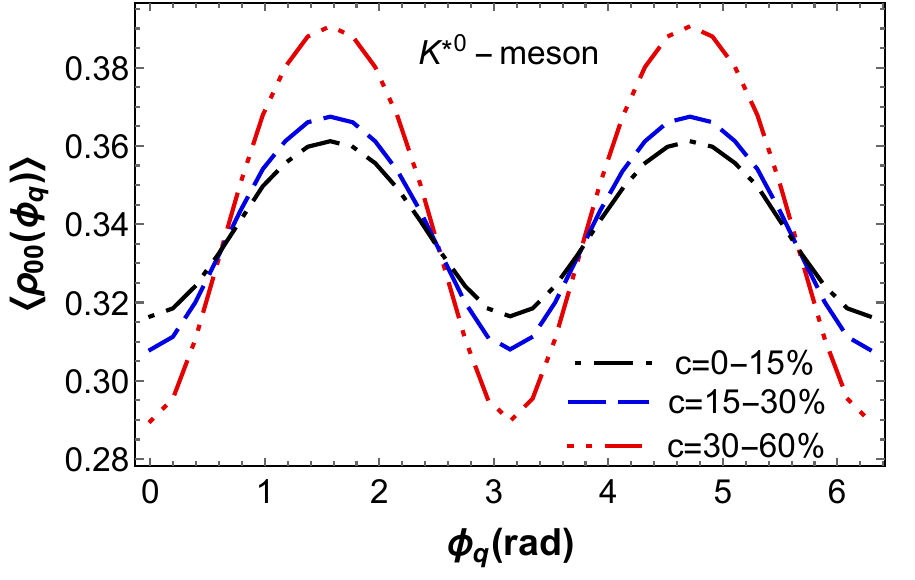}
        {}\includegraphics[width=0.45\textwidth]{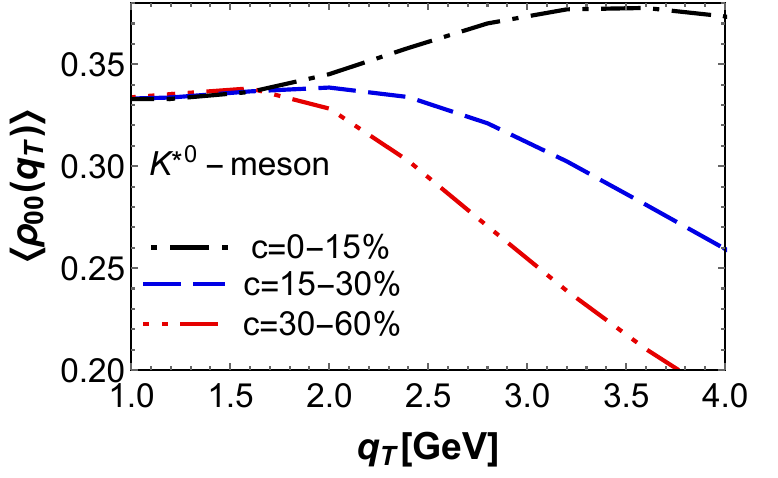}
        {}\includegraphics[width=0.45\textwidth]{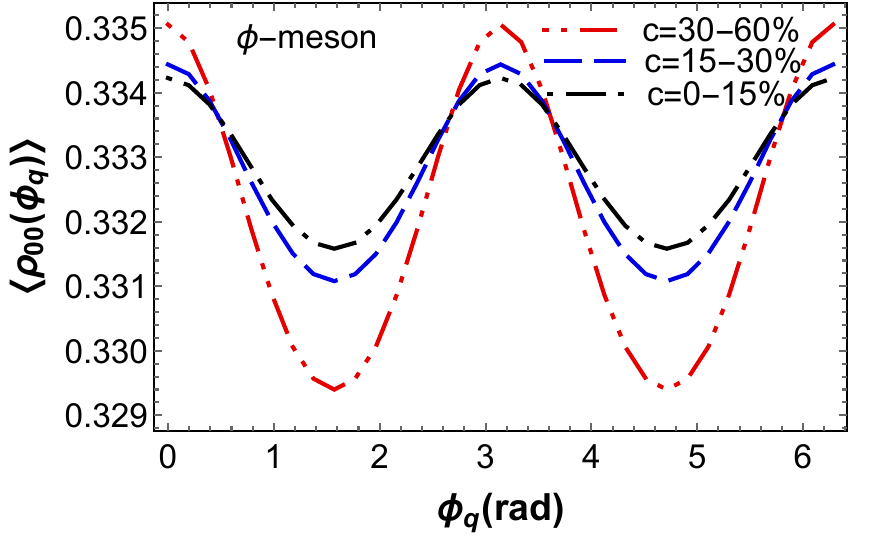}
        {}\includegraphics[width=0.45\textwidth]{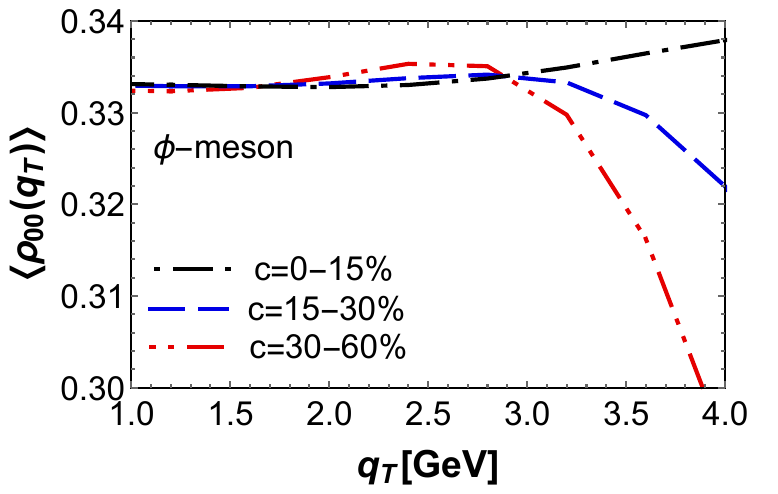}
		\caption{Azimuthal angle (left panel) and transverse momentum dependence (right panel) of the $\rho_{00}$ component of the spin density matrix for $K^{*0}$ (upper panels) and $\phi$ mesons (lower panels) for the centrality classes $c$=0--15\%, $c$=15--30\% and $c$=30--60\% with freeze-out temperature $T_f=165$ MeV and collision energy $\sqrt{s_{NN}}=130$ GeV at $y_p=0$.
		}
	 \label{fig:f2}
\end{figure*}

In FIG.~\ref{fig:f2}, we present the numerical results for the azimuthal angle ($\phi_q$)  (left panel) and the transverse momentum ($q_T$) dependence 
(right panel) of $\rho_{00}$ as defined in Eqs.~(\ref{eq:rho_phi}-\ref{eq:P_pT}) for $K^{*0}$ and $\phi$. The azimuthal angle dependence is obtained by integrating over $q_T$ in the range $1.2-5.4$ GeV. The corresponding plots indicate that $\langle\rho_{00}(\phi_q)\rangle$ can be accurately parametrized as  
$\langle\rho_{00}(\phi_q)\rangle = a \cos(2\phi_q) + b$ with $b\simeq 1/3$, while $a<0$ for $K^{*0}$ and $a>0$ for $\phi$. Our estimate for the 
values of the coefficients $a$ and $b$ obtained  by fitting the above defined function $\langle\rho_{00}(\phi_q)\rangle$ to the  numerical data for three different centrality classes are listed in  the Table~\ref{tab1}. 
Such values can be probed in future RHIC and LHC experiments. The sign change of $a$ for $K^{*0}$ and $\phi$ might be inferred from their dependence on $q_{x,y}$ for $\delta \rho_{00}$. As shown in FIG.~\ref{fig:f1}, for $q_y=0$ and equivalently $\phi_q=0$, one finds that $\delta \rho_{00}$ is mostly negative for $K^{*0}$ despite being positive in a marginal region for $\bar{q}_x \sim 0$, where $\bar{q}_{x}=|q_x|-1.2$ GeV,  
while for $\phi$, $\delta \rho_{00}$ is mostly positive in a larger region around $\bar{q}_x \sim 0$, which is more prominently weighted compared to the large-$\bar{q}_x$ region. When integrating over $q_T$ (more precisely $\bar{q}_x$), it turns out that $\delta \rho_{00}$ at $\phi_q=0$ may thus become negative and positive for $K^{*0}$ and $\phi$, respectively. Analogously, one finds the opposite patterns for $\phi_q=\pi/2$. We hence observe the opposite oscillatory pattern of $\langle\rho_{00}(\phi_q)\rangle$ for $K^{*0}$ and $\phi$. Since $\mathcal{N}$ is suppressed by larger $m$, the overall deviation of $\langle\rho_{00}(\phi_q)\rangle$ from $1/3$ (if nonzero) is more prominent for $K^{*0}$ as shown in FIG.~\ref{fig:f2}. Overall, our finding shows that the azimuthal-angle dependence of $\rho_{00}$ is rather sensitive to the mass values of the considered vector mesons.
\begin{table*}[ht!]
\centering
\begin{tabular}{ |p{3cm}||p{3cm}||p{3cm}||p{3cm}||p{3cm}|} 
   \hline
 c $\%$ & $a$ ($K^{*0}$--meson)&   $a$ ($\phi$--meson) &   $b$ ($K^{*0}$--meson) &   $b$ ($\phi$--meson)\\
  \hline
   $0-15$ &   $-0.0223$ &   $0.0013$ &   $0.3399$ &   $0.3329$ \\
  $15-30$ &   $-0.0295$ &   $0.0017$ &   $0.3406$ &   $0.3327$ \\
  $30-60$ &   $-0.0498$ &   $0.0028$ &   $0.3453$ &   $0.3321$ \\
 \hline
\end{tabular}
\caption{Values of the coefficients $a$ and $b$ obtained by fitting the function $\langle\rho_{00}(\phi_q)\rangle = a \cos(2\phi_q) + b$ 
to our numerical data for different centrality classes.}
\label{tab1}
\end{table*}

Our results of $\langle\rho_{00}(q_{T})\rangle$ for $K^{*0}$ and $\phi$ indicate that in different momentum ranges the values to $\rho_{00}$ can be  $\leq1/3$ as well as $\geq1/3$. At small transverse momentum ($q_T<1.7$ GeV) one notices that for both particles $\rho_{00}$ is roughly $1/3$ for all the three centrality classes. At higher momenta of the order of $q_T\sim 4$ GeV, $\rho_{00}$ can be larger than 1/3 for the lowest centrality class of $c=0-15\%$, but decreases considerably below 1/3 with increasing $q_T$ for higher centralities. 

\begin{figure}[ht!]
      \centering
        {}\includegraphics[width=0.45\textwidth]{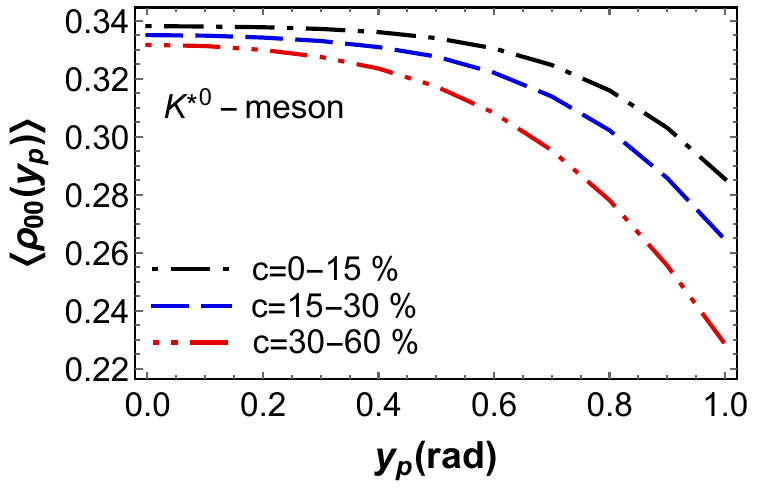}
    {}\includegraphics[width=0.45\textwidth]{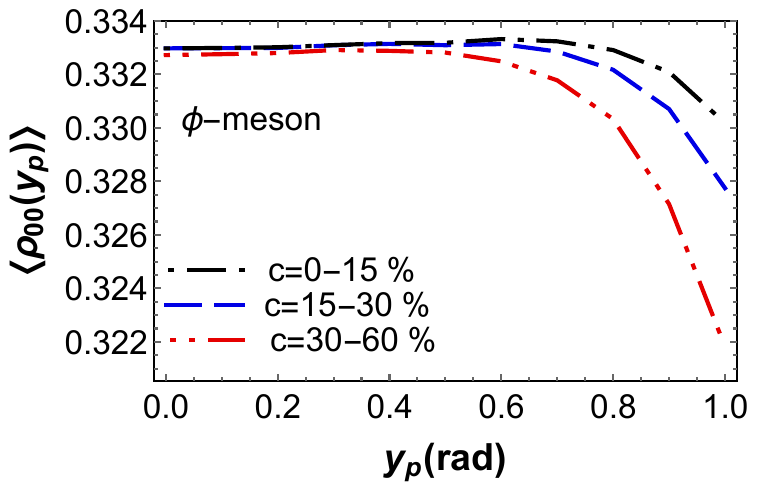}
		\caption{Rapidity dependence of the $\rho_{00}$ component of the spin density matrix for $K^{*0}$ (left panel) and $\phi$ mesons (right panel) for the centrality classes $c$=0--15\%,  $c$=15--30\% and $c$=30--60\% with freeze-out temperature $T_f=165$ MeV and collision energy $\sqrt{s_{NN}}=130$ GeV.
		}
	 \label{fig:f3}
\end{figure}

In FIG.~\ref{fig:f3}, we depict the rapidity dependence of $\rho_{00}$ for $K^{*0}$ and $\phi$. In the case of $K^{*0}$, one notes that $\rho_{00}$ decreases with increasing rapidity $y_p$ and centrality $c$. In contrast, $\rho_{00}$ for the $\phi$ initially increases with respect to $y_p$ then decreases rapidly below 1/3 at larger  $y_p$ for all centrality classes with more prominent decrease at higher centrality.
Since $\rho_{00}$ for the $\phi$ is close to $1/3$ in the region of $|y_p| \lesssim 1$, it can be expected that its global value by further integrating over $y_p$ is also close to $1/3$. For the global spin alignment of $K^{*0}$, the deviation from $1/3$ may be larger than that of $\phi$, but still rather small.

\section{Summary and Outlook}\label{sec:summary_outlook}
In this paper, expanding the Wigner function of vector mesons up to $O(\hbar^2)$, we have derived an expression for $\rho_{00}$ in terms of their distribution functions. 
We further discussed the large-momentum and small-momentum limits of $\rho_{00}$. 
As a result, we find that the second-order space-time gradients of the vector-meson distribution functions can trigger spin alignment, 
leading to a deviation of $\rho_{00}$ from 1/3. Next, by considering  the J{\"u}ttner type equilibrium distribution and using 
a thermal model with single freeze-out, we computed several spin alignment observables for Au-Au collisions with $\sqrt{s_{NN}}= 130$ GeV. 
Studying the dependence of $\rho_{00}$ on the azimuthal angle $\phi_q$ at midrapidity, we found a $\langle \rho_{00}(\phi_q)\rangle = a \cos(2\phi_q) + b$ pattern, with $b \simeq 1/3$ for both $K^{*0}$ and $\phi$, but $a<0$ for $K^{*0}$ and $a>0$ for $\phi$. 
For $\langle \rho_{00}(q_T)\rangle$ at midrapidity, our results indicate that the deviations of $\rho_{00}$ from 1/3 could be greatly enhanced at large transverse momenta for different collision centralities. We furthermore studied the rapidity dependence of $\rho_{00}$ for $K^{*0}$ and $\phi$ at different centralities, 
obtaining a rather different behavior for the two particles, as shown in Fig.\,\ref{fig:f3}.
For more accurate and versatile estimates, that go beyond the simple model used in this work, we will in the future 
conduct hydrodynamic simulations with broader ranges of collision energies.

As mentioned in the discussion of our setup in Section \ref{Sec:wf}, we have neglected dynamical contributions to $\rho_{00}$ 
from the polarization dependent $\tilde{f}_{\lambda}(\bm q,X)$. For example, early-time effects could exist, which lead to further 
$\mathcal{O}(\hbar^2)$ corrections to $\tilde{f}_{\lambda}(\bm q,X)$, such as glasma fields \cite{Kumar:2022ylt, Kumar:2023ghs}, turbulent color fields in the QGP phase \cite{Muller:2021hpe,
Yang:2021fea}, and  fluctuating meson fields before hadronization \cite{Sheng:2019kmk,Sheng:2020ghv,Sheng:2023urn}, which interact with the constituent quark and antiquark forming the vector meson in the quark coalescence scenario \cite{Fries:2003vb,Greco:2003xt}. 
In addition, in the hadron phase, polarization dependent interactions could further modify $\tilde{f}_{\lambda}(\bm q,X)$, which 
have to be treated within the quantum kinetic theory of vector-mesons with proper collision terms as noted in Section \ref{Sec:wf}. 
All these dynamical contributions should be combined with the non-dynamical contribution led by the second-order gradients upon $\tilde{f}_{\lambda}(\bm q,X)$ with a proper choice of the freeze-out hypersurface found in this work to address experimental observations. 

On the other hand, since the effect we considered is only related to the final state of vector-meson distribution functions, it is independent of detailed mechanisms for hadronization such as the quark coalescence or fragmentation. As a result, Eq.~(\ref{eq:rho00q}) is applicable to all types of vector mesons with arbitrary momenta and not subject to just $K^{*0}$ and $\phi$. In the future, given $\tilde{f}_{\lambda}(\bm q,X)$ at certain freeze-out points obtained from prescribed transport models, 
it will become possible to study this effect 
for $J/\psi$ and $D^{*+}$ at the LHC or $K^{*0}$ and $\phi$ in low-energy nuclear collisions.
\label{sec:sum}
\begin{acknowledgments}
D.Y. was supported in part by the National Science and Technology Council (Taiwan) under Grant No. MOST 110-2112-M-001-070-MY3. 
P. G. is supported by KAKENHI under Contract No. JP20K03940, JP20K03959 and JP22H00122.
\end{acknowledgments}

\newpage
\appendix
\section{Different components of $\Pi^{(\mu\nu)}(0, q)=\epsilon^{(\mu}(0,q)\epsilon^{*\nu)}(0,q)$}
We start by writing Eq.(\ref{pol_vec1}) as 
\begin{eqnarray}
\epsilon^{\mu}(\lambda, q)&=&\Big(\frac{-q_{\alpha}\epsilon^{\alpha}_{{\lambda},\perp}}{M},\epsilon^{\mu}_{{\lambda},\perp}-\frac{q_{\alpha}\epsilon^{\alpha}_{{\lambda},\perp}}{M(E_q+M)}q^{\mu}_{\perp}\Big). \quad\quad\quad \label{pol_vec1_cov}
\end{eqnarray}
The spin vectors ${\bm\epsilon^i_{\lambda,\perp}}={\bm\epsilon^i_{\lambda}}$ appearing in the  above equation are given by Eqs.(\ref{e0}-\ref{e-1}) and satisfy the condition  Eq.(\ref{eps_r2}) and ${\epsilon^0_{\lambda,\perp}}=0$. 

Using Eq.(\ref{pol_vec1_cov}) along with Eqs.(\ref{e0}) one obtains
\begin{eqnarray}
\Pi^{(00)}(0, q)=\epsilon^{(0}(0,q)\epsilon^{*0)}(0,q)&=&\frac{2(q^y)^2}{M^2}
\quad\quad\quad\quad\label{eq:eps00}
 \end{eqnarray}
 Similarly, using Eq.(\ref{pol_vec1_cov}) with Eqs.(\ref{e0}-\ref{e-1}) on finds 
 \begin{eqnarray}
\Pi^{(0i)}(0, q)=\epsilon^{(0}(0,q)\epsilon^{*i)}(0,q)
&=&\frac{2 q^y}{M}\left(\epsilon^{i}_{0}+\frac{q^y q^i}{M(E_q+M)}\right),
  \quad\quad\quad\quad\label{eq:eps0i1}
 \end{eqnarray}
 and
 \begin{eqnarray}
\Pi^{(ij)}(0, q)=\epsilon^{(i}(0,q)\epsilon^{*j)}(0,q)
&=&2\bm\epsilon^{i}_{0} \epsilon^{j}_{0}
+\frac{2 q^y\left(q^i \epsilon^{j}_{0}
+q^j \epsilon^{i}_{0}\right)}{M(E_q+M)} +\frac{2 (q^y)^2 q^i q^j}{M^2(E_q+M)^2}.
\quad\quad\quad\quad~~~~~~~~~~~\label{eq:eps0ij1} 
\end{eqnarray}
%
%
\section{Different components of $\Pi^{(\mu\nu)\alpha\beta}_{(2)}(q)$}\label{app:components_Pi2}
The tensor, $\Pi^{(\mu\nu)\alpha\beta}_{(2)}(q)$ is defined as
\begin{eqnarray}
\Pi^{(\mu\nu)\alpha\beta}_{(2)}(\lambda,q)\equiv \Bigg(\left(\partial_{q}^{{\alpha}}\partial_{q}^{{\beta}}\epsilon^{(\mu}(\lambda,q)\right)\epsilon^{*\nu)}(\lambda,q)-\big(\big(\partial_{q}^{{\alpha}}\epsilon^{(\mu}(\lambda,q)\big)\big(\partial_{q}^{{\beta}}\epsilon^{*\nu)}(\lambda,q)\big)\big)\Bigg)+{\rm c.c.} \quad \label{Pimnabsymrel1.1}
\end{eqnarray}
Using Eq.(\ref{pol_vec1_cov}) on can easily show
\begin{eqnarray}
\Pi_{(2)}^{(00)\alpha\beta}(q)=A_1\eta^{\alpha\beta},
\end{eqnarray}
with
\begin{eqnarray}
A_1=\frac{2}{M^2},
\end{eqnarray}
where the factor 2 comes from the complex conjugate. 

Now, using Eqs.(\ref{pol_vec1_cov}) and (\ref{eps_r2}) one can write
\begin{eqnarray}
\Pi_{(2)}^{(0i)\alpha\beta}(q)&=&A_2\eta^{\alpha i}q^{\beta}+B_2\eta^{\alpha \beta} q^{i}+C_2q^{\alpha}\eta^{\beta i}+D_2q^{\alpha}q^{i}q^{\beta},
\end{eqnarray}
where
\begin{eqnarray}
A_2&=&-2\frac{1}{M^2(E_q+M)}\left(1+\frac{\bm|q|^2}{E_q(E_q+M)}\right)=-\frac{2(2E_q-M)}{M^2E_q(E_q+M)},\nonumber\\
B_2&=&2\frac{1}{M^2(E_q+M)}\left(2-\frac{\bm|q|^2}{E_q(E_q+M)}\right)=+\frac{2}{M^2E_q},\nonumber\\
C_2&=&\frac{2}{M^2(E_q+M)},\nonumber\\
D_2&=&2\frac{1}{M^2E_q(E_q+M)^2}\left(\frac{(3E_q+M)\bm|q|^2}{E^2_q(E_q+M)}\right)=\frac{2(3E_q+M)(E_q-M)}{M^2E^3_q(E_q+M)^2}.\label{eq:coef1}
\end{eqnarray}
Similarly, one can write
\begin{eqnarray}
\Pi_{(2)}^{(ij)\alpha\beta}(q)
&=&
\bigg[A_3\eta^{\alpha (i}\eta^{\beta j)}+B_3q^{(i}\eta^{\beta j)}q^{\alpha}+C_3q^{(i}\eta^{\alpha j)}q^{\beta}+D_3\eta^{\alpha\beta}q^{(i}q^{j)}+E_3q^{\alpha}q^{\beta}q^{(i}q^{j)}\bigg]
\quad\label{Pi_ijab2}
\end{eqnarray}
where
\begin{eqnarray}
    A_3&=&2\frac{1}{M(E_q+M)}\left(2-\frac{|\bm q|^2}{M(E_q+M)}\right)=2\frac{1}{M(E_q+M)}\left(3-\frac{E_q}{M}\right),\nonumber\\
    B_3&=&-2\frac{1}{ME_q(E_q+M)^2}\left(1-\frac{|\bm q|^2}{M(E_q+M)}\right)=-2\frac{1}{ME_q(E_q+M)^2}\left(2-\frac{E_q}{M}\right),\nonumber\\
    C_3&=&-2\frac{1}{M(E_q+M)^2}\left(\frac{2}{E_q}\right),\nonumber\\
    D_3&=&2\frac{1}{M(E_q+M)^2}\left(\frac{1}{M}-\frac{1}{E_q}\left(1+\frac{|\bm q|^2}{M(E_q+M)}\right)\right)=0,\nonumber\\
    E_3&=&2\frac{1}{ME_q(E_q+M)^3}\left(\frac{(3E_q+M)}{E^2_q}+\frac{1}{M}\left(2+\frac{(3E_q+M)|\bm q|^2}{E^2_q(E_q+M)}\right)-\frac{1}{M}\left(2+\frac{|\bm q|^2}{E_q(E_q+M)}\right)\right)\nonumber\\&&={\frac{4}{M^2E^2_q(E_q+M)^2}}.\nonumber\\
\end{eqnarray}
%
%
%

\bibliographystyle{utphys}
\bibliography{polarization_ref}

\end{document}